\documentclass[12pt,preprint]{aastex}
\usepackage{epsfig}

\shorttitle{Serpens NW}
\shortauthors{Hodapp,Chini,Watermann,Lemke}

\begin{document}

\title{Eruptive Variable Stars and Outflows in Serpens NW}
\author{Klaus W. Hodapp\altaffilmark{1},
Rolf Chini\altaffilmark{2}, 
Ramon Watermann\altaffilmark{2}, 
Roland Lemke\altaffilmark{2}, 
}

\altaffiltext{1}{
Institute for Astronomy, University of Hawaii,
640 N. Aohoku Place, Hilo, HI 96720, USA
\\email: hodapp@ifa.hawaii.edu }

\altaffiltext{2}{
Ruhr Universit{\"a}t Bochum, Astronomisches Institut,
Universit{\"a}tsstrasse 150, D-44801 Bochum, Germany\\
}

\begin{abstract} 

We study the outflow activity, photometric variability and morphology
of three very young stellar objects in the Serpens NW star forming region: OO~Serpentis,
EC~37 (V370~Ser) and EC~53 (V371~Ser). 
High spatial resolution Keck/NIRC2 laser guide star adaptive optics images obtained
in 2007 and 2009 in broad-band \textit{K} and in a narrow-band filter centered 
on the 1--0 S(1) emission line of H$_2$ allow us
to identify the outflows from all three objects. 
We also present new, seeing-limited data on the photometric evolution of the
OO~Ser reflection nebula and re-analyze previously published data.
We find that OO~Ser declined in brightness from its outburst peak in 1995 to about 2003, but that this
decline has recently stopped and actually reversed itself in some areas
of the reflection nebula.
The morphology and proper motions of the shock fronts MHO 2218 near EC~37 
suggest that they all originate in EC~37 and that this is an outflow seen nearly along
its axis. 
We identify an H$_2$ jet emerging from the cometary nebula EC~53. 
The star illuminating EC~53 is periodically variable with a period of 543 days,
and has a close-by, non-variable companion at a projected distance of 92 AU.
We argue that the periodic variability is the result of accretion instabilities
triggered by another very close, not directly observable, binary companion and
that EC~53 can be understood in the model of a multiple system developing into
a hierarchical configuration.

\end{abstract}

\keywords{
binaries: close ---
ISM: jets and outflows ---
stars: formation --- 
stars: low-mass ---
stars: variables: T Tauri, Herbig Ae/Be
}

\section{INTRODUCTION}

The Serpens star forming region is a compact region of low-mass star
formation.
At a distance of 311 pc \citep{deLara1991} it is still close
enough for detailed, spatially resolved studies of its individual young
stellar objects (YSOs) and has therefore attracted considerable attention.
The Serpens NW region is part of the Serpens molecular cloud and is the site
of much of the most recent star formation activity in this cloud, 
as indicated by multiple sources of CO outflow and
H$_2$ jets \citep{Hodapp1996, Herbst1997} in an area of less than 0.2 pc in projected diameter.
After two decades of monitoring this region in the \textit{K} band, it is clear that
two objects in Serpens NW are particularly noteworthy in
the context of establishing the relationship between outflows, 
accretion instabilities, and multiplicity: 
Serpens DEOS = OO~Ser, and EC~53 = V371~Ser.

The Serpens NW molecular core contains the unique eruptive variable
OO~Ser discovered by \citet{Hodapp1996} and originally named the ''Deeply Embedded
Outburst Star (DEOS)''. This object has
an outburst amplitude similar to the larger and more luminous FU Orionis (FUor)-class
of objects and an outburst duration longer than that of EX Lupi (EXor) objects.
(See \citet{Herbig1977} for the definition of these object classes.)
It is so deeply embedded as to be inaccessible at optical wavelengths and
shows an infrared spectrum \citep{Hodapp1996, Hodapp1999, Kospal2007} 
much more dominated by dust than either of those
established classes of young eruptive variables. 
Based on its association with a bipolar nebula, and its detection at
sub-mm and millimeter wavelengths, \citet{Hodapp1996} have classified this
object as Class I in the scheme of \citet{Adams1987}. 
This conclusion was corroborated by
\citet{Kospal2007} who found, on the basis
of archival Spitzer data, that OO~Ser is the dominant source in Serpens NW at mid-infrared wavelengths.
This point was further strengthened by \citet{Evans2009}
who observed the Serpens star forming region with Spitzer as part of the ''Clouds to Disks'' (C2D) project. Among the sources
in Serpens NW classified in their survey, OO Ser has the lowest 
bolometric temperature (130 K) and highest luminosity (10 $L_{\sun}$).
OO~Ser consists of an extended reflection nebula and a more compact
central flux maximum. The latter appears slightly extended on our adaptive optics
images with a cometary morphology opening to the west 
and therefore represents the most intense inner parts of the
reflection nebula. At near-infrared wavelengths, we probably do not see any
direct, unscattered light from the embedded star. We therefore use the variable star designation
OO~Ser to describe the variable reflection nebula in its entirety.

The variable object V371 Ser, originally named EC~53 in the IR imaging survey of
the Serpens molecular cloud by \citet{Eiroa1992},
and the irregularly variable star V370 Ser (EC~37)
have also attracted attention.
Both these sources are probably among the more evolved YSOs in the 
Serpens NW region, since they are quite bright at near-infrared wavelengths
and do not dominate at mid-IR wavelengths.
V371 Ser (EC~53) is a "cometary" nebula, i.e., one visible lobe of a bipolar
structure. 
Both objects are associated with molecular hydrogen shock fronts indicating
ongoing outflow and therefore accretion activity.

One goal of this study was to determine the source of the 
faint H$_2$ v=1-0 S(1) emission labeled MHO 2218 by \citet{Davis2010}
in the area of EC~37 and OO~Ser, to determine which of these 
two objects is the driving source of this shock excited emission. 
In a complex, dense star-forming region such a Serpens NW, 
it is not trivial to properly associate H$_2$ shock fronts with 
their driving sources, yet such proper identification is
essential for determining the evolutionary state of the YSOs.
Another purpose was to study the bipolar nebula associated
with the eruptive variable object OO~Ser and the changes 
it underwent since its last observed eruption in 1995. 
The cometary nebula EC~53 was fortuitously included in the field of view
and the study of this object returned very interesting results.

We have observed the Serpens NW region with the Keck II
laser adaptive optics system
both in broad band \textit{K$_s$} and in H$_2$ line emission on two occasions separated by
about two years. The observations are listed in Table 1, described in detail in section 2.1, and
the results are discussed in section 3.1.
To further investigate the temporal evolution of the light distribution in the
OO Ser bipolar nebula, older, seeing-limited data, primarily from the UH 2.2 m
telescope, but also recently obtained data
from the Infrared Imaging System (IRIS), UKIRT, and the CFHT were analyzed in addition to the Keck data.
These observations are described in Section 2.2, 
are listed in Table 2, and discussed in Sections 3.2 and 3.3.

\section{OBSERVATIONS AND DATA REDUCTION}

\subsection{Keck Adaptive Optics Imaging}
The adaptive optics data reported here were obtained at the Keck II telescope, using
the laser guide star adaptive optics (LGSAO) system \citep{Wizinowich2006}
in conjunction with the Near-Infrared Camera 2 (NIRC2) camera. 
The dates, details of the observations, and the achieved FWHM image quality of the
final images are summarized in Table 1.
In addition to the object images, we took separate sky frames at positions that
appeared free of bright stars and extended emission on deep, seeing-limited images taken
earlier. The dithered sky frames were also adaptive optics corrected 
to allow the optimal removal 
of any faint stars
by computing the median of a set of such sky frames.
Flatfield correction was done using differential (lights-on minus lights-off)
dome flats.
The Serpens NW region is a dense nearby molecular cloud and consequently suffers from
a lack of available optically visible tip-tilt guide stars. Within the constraints of the Keck
LGSAO system, the only available guide star is EC~92 (2MASSJ18295360+0117017) with an R magnitude of 16.6. 
Even with this tip-tilt guide star, OO~Ser can only be reached when using 
the wide field (40$\arcsec$ FOV) mode of
NIRC2, and only with a rotator position of $\approx$279$\arcdeg$ to 
place the tip-tilt star in the extreme corner of the guider field. 
These circumstances explain the otherwise
strange choice of field orientation and object placement in our images.
It is obvious that the object placement is less than ideal, 
with OO~Ser in one corner of the 
field, and EC~53 in the bad quadrant of the NIRC2 detector, where
one of the 8 output amplifiers of the Aladdin detector
array is not operable and several others shows excess noise.
The images were processed using various IRAF routines (see Section 2.2 for a general
description of the data reduction procedure) and were astrometrically
de-warped using the IDL routines provided by B. Cameron on the NIRC2 website
\footnote{http://www2.keck.hawaii.edu/inst/nirc2/forReDoc/post\_observing/dewarp/}.

We have obtained proper motion measurements of the various H$_2$ shock fronts
from the H$_2$ S(1) narrow-band images obtained in 2007 and 2009, because these
images had the best signal-to-noise ratio for the faint shock fronts
yet still showed a sufficient number of stars to use as an astrometric
reference. By comparison, the \textit{K}-band images had more S/N on the stars,
but the shock fronts were generally too faint to get good astrometric
results. Establishing an astrometric reference grid on a small image
near the center of a nearby molecular cloud is fraught with difficulties
and has to rely on some simplifying assumptions. There are no stars in
this image (Fig. 1) that could be confidently identified as distant background
stars with negligible proper motion. The few stars in the image could
possibly be highly reddened background stars, but they could also be
faint, partly embedded young stars within the Serpens NW star-forming
region. The best that can be done in this situation is to use all available
stars as an astrometric reference. The images at the two epochs
were astrometrically matched using the IRAF tasks \textit{geomap} and \textit{geotran}. 
Since the images had already been de-warped, we restricted the \textit{geomap} solution to
linear terms, to avoid instabilities due to the small number of reference stars.
Proper motions were then measured using the IRAF task \textit{xregister} 
that computes the cross-correlation between the
images at the two epochs in a
sub-image containing a feature (star or shock front) of interest.

We summarize our proper
motion measurements in Fig. 1, where the measurements of stars are 
represented by amber (light grey) arrows, while those of H$_2$ shock fronts are
indicated by blue (dark grey) arrows. 
All the stars have small residual proper
motions against the reference system established by the ensemble of
these same stars, and these proper motion vectors represent the combined
error of the individual proper motion measurements and the uncertainty of the
astrometric reference system defined by that ensemble of stars, as well
as possibly genuine individual proper motion.

\subsection{Photometric Monitoring}
The study of the temporal evolution of the reflection nebula 
associated with OO~Ser and the variability of EC~53
made use of older observations obtained at the UH 2.2 m telescope 
with the QUIRC camera \citep{Hodapp1996QUIRC} that
had already been published in the earlier papers on 
the discovery of OO~Ser \citep{Hodapp1996} and the initial
study of the OO~Ser and EC~53 light curves \citep{Hodapp1999}. 
To these light curves, we have now added
newer QUIRC data obtained up to 2004 when that camera was decommissioned, 
data obtained at the UH 2.2 m telescope with the SIRIUS
camera \citep{Nakajima2002} in 2000,
as well as data obtained at UKIRT with WFCAM in 2008 and at CFHT with WIRCAM in 2011.
We have also included
new data from the recently commissioned 0.8 m Infra-Red Imaging Survey (IRIS) telescope 
described by \citet{Hodapp2010}.
Finally, we have used the oldest available near-infrared
image of the Serpens NW region that we had obtained on
1991, Aug. 5 with the UH NICMOS-3 camera \citep{Hodapp1992} at the UH 2.2 m
telescope in the \textit{K$^\prime$} filter \citep{Wainscoat1992}. 
We used the same photometric apertures
as for the other data sets and the same reference stars.
The correction between \textit{K$^\prime$} and \textit{K} was determined
from near-simultaneous images obtained in those two filters during the
outburst of OO~Ser in 1995.
The lowest spatial resolution data, the
1991 UH NICMOS3 camera image and the 2010 IRIS 0.8 m telescope
data, were not used for the sub-regions in the reflection
nebula.
The seeing-limited observations are listed in Table 2.

For the study of the photometric behavior of individual knots 
of nebulosity within the OO~Ser bipolar nebula, 
we have also included the two Keck \textit{K}-band images (Table 1), after scaling
them to the common astrometric solution 
of the seeing-limited images.
The Keck images could not be used to measure the integral brightness
of the reflection nebula, because their field of view did not contain
the integration box for the integral brightness completely.

The images were flatfield corrected using differential dome
flats, bad pixel masks were derived from the dome flats.
The dithered object images were median combined into 
sky frames and subtracted from the flatfielded images.
The resulting images were aligned and co-added while
eliminating bad pixels in the process.
After this standard data reduction, these seeing-limited images were 
transformed to a common astrometric solution using the
IRAF tasks \textit{geomap} and \textit{geotran}. 
The available data were diverse in the total exposure time achieved,
prevailing seeing, and photometric quality of the nights. 
For most, separate standard star observations were not
available. Therefore, the photometric calibration relied
on an ensemble of field stars. Of these field stars, we
selected the four stars that showed the smallest scatter
of the photometric measurements in the 1994 to 1998 time
interval as secondary standards and calibrated them against
brighter stars in the 2MASS catalog \citep{skr06}. 
Obviously, our procedure makes the assumption that these
selected stars were constant throughout the two decades for
which photometry is being discussed here.
We have not tried
to correct for color terms between the \textit{K$_s$} filter of 2MASS \citep{Cohen2003},
the classical \citep{Johnson1966} \textit{K}, and the Mauna-Kea-standard \textit{K} \citep{Tokunaga2002}
filters used in the UH QUIRC camera and the other cameras occasionally used, because such
effects are much smaller than the variability amplitudes discussed here.
To avoid problems
caused by the ubiquitous extended emission, 
photometry on rectangular sub regions
of the bipolar nebula associated with OO~Ser were obtained 
with the IRAF task \textit{imstat}.
The sky background was separately measured in similar rectangular 
regions selected to be unaffected by stars or extended flux.

In a diverse data set combining data taken with different telescopes
and cameras, different exposure times and under varying conditions,
the internal error of a photometric measurement underestimates
the uncertainty of the result. In our case, for the higher quality
data obtained with QUIRC at the UH 2.2 m telescope, internal errors
are only of the order of $10^{-5}$ Jy. 
For the UH 2.2 m telescope QUIRC
and SIRIUS data, the rms scatter of the individual secondary 
standard star flux values over time is $\approx3\times10^{-5}$ Jy. 
The IRIS 0.8 m telescope data have higher rms scatter of $\approx5\times10^{-5}$ Jy, due to
poorer sampling and smaller signal-to-noise ratio.
With these total errors, the variations in the light curve for different regions
of the OO~Ser reflection nebula are highly significant.

\section{Results and Discussion}

\subsection{Proper Motion of H$_2$ Shock Fronts Near EC~37 = V370 Ser}

Figure~1 shows the proper motion vectors measured from the Keck NIRC2 H$_2$ S(1)
adaptive optics images taken in 2007 and 2009. 
For many of the clearly defined bow shock systems, these results
essentially confirm the earlier results by \citet{Hodapp1999} that
were based on seeing-limited images over a 4 year interval.
The new data with their superb spatial resolution and
depth enable, for the first time, a study of
the much fainter emission in the general area of EC~37, 
which is listed as MHO2218 in the catalog of H$_2$ shock fronts by \citet{Davis2010}
\footnote{The catalog is available at http://www.jach.hawaii.edu/UKIRT/MHCat.}
Figure~2 is a version of the 2009 Keck AO image where an intensity-scaled version
of the \textit{K} image was subtracted, to best bring out the H$_2$ shock fronts by
subtracting out the continuum sources.

The young stellar object EC~37 has an average \textit{H-K} $\approx$3.6
from the photometry reported by \citet{Hodapp1996}, making it one of the
reddest of the point sources in the catalog of Serpens infrared sources by \citet{Eiroa1992}. 
As already shown by \citet{Hodapp1999} it exhibits the irregular variability characteristics of 
T~Tauri stars \citep{Joy1945}.
In addition, it coincides with a small unresolved knot 
of 3 mm emission at flux level 3.6 mJy in the
map of \citet{Testi1998}, indicating the presence of substantial amounts
of cold dust around the star, and therefore a young age of the object.
On the adaptive optics H$_2$ 1--0 S(1) images, individual knots of emission
are visible in the immediate vicinity of EC~37, primarily to the south
of the star, in addition to more smoothly distributed flux. 
Very faint, knotty emission can be seen up to about 5$\arcsec$ to the
south of EC~37. Outside of this area, the brightest H$_2$ shocks 
are found to the east and north of EC~37.

The large system of
H$_2$ S(1) shock fronts near EC~37 (MHO 2218) shows a pattern of proper motion vectors
that can best be understood as an outflow emerging from EC~37.
The emission knots east of EC~37 were already imaged 
and their proper motion measured by
\citet{Hodapp1999}.
For those brighter knots, the older 
measurements agree very well with the new adaptive-optics-based  
proper motion vectors.
In addition, fainter emission knots to the south-east of EC~37
could now be measured that show unambiguously that those knots
move away from EC~37 and not from OO~Ser, which is located even further south. 
A few of the small emission
knots in the immediate vicinity of the unresolved star EC~37 also
have measured proper motion now, showing a pattern of motion radially
away from the star. 

This system of emission knots near EC~37 is clearly not seen in an orientation close
to the edge-on orientation that is favorable for seeing the full
extent of a jet. The comparatively low extinction towards the
central star that makes it directly visible at near-infrared wavelengths, 
the fact that H$_2$ S(1) emission knots are seen all
around EC~37, and the complex pattern of proper motion vectors
all support the notion that
this outflow is seen nearly along the direction of its axis.
This is also supported
by the general lack of morphological features that would be indicative of
an edge-on orientation, such as a cometary or bipolar morphology.
In Serpens NW, both EC~53 and
OO Ser are examples of the latter morphology.

\subsection{OO Ser}

The deeply embedded eruptive variable source OO~Ser shows 
nebulosity of bipolar morphology, 
with the western lobe being brighter than the eastern. 
At the spatial resolution of our adaptive optics
co-added images of $\approx$100 mas FWHM (see Table 1), the object at the center 
of the bipolar nebula (in the south-western corner of Fig. 1) 
appears slightly extended and elongated with the
shape of a cometary nebula opening to the west. Even on these deep 
adaptive optics images, we are not seeing direct light from the central
star of the OO~Ser reflection nebula. 
The whole reflection nebula brightened after the outburst of the central object
in 1995. However, the changes in brightness were not spatially uniform.
Motivated by these tentative findings, we have re-analyzed older, seeing-limited images
from the time of maximum brightness in 1995 to images taken in 2008. The
images were spatially registered to the image taken near maximum brightness 
in July of 1995 using the IRAF tasks \textit{geomap} and \textit{geotran}. 
After adjusting the sky background to zero
and after scaling of the images by the integrated flux of the reflection nebula 
to account for the varying average brightness of the
extended object, difference images were computed. These are shown in Fig.~3 and
illustrate that the bipolar nebula did not change its brightness spatially uniformly.
In Fig.~3, a faster than average decline in brightness is represented by dark
tone while areas declining more slowly appear bright.
Temporal variations in the flux distribution of reflection nebulae are
not a new phenomenon. For example, the case of R Mon has been studied as
early as \citet{Hubble1916}. Examples observed in the near-infrared 
include Cep~A \citep{Hodapp2009} and L483 \citep{Connelley2009}. 
The interesting aspect of these variations is that they point to
significant variations of the extinction in the optical path of the
scattered light on time scales of a few years, making the spatial scales 
of the obscuring dust comparable to those of our own Solar System. 

We have also extracted photometry of the OO~Ser nebula and its central extended source
on the available seeing limited images, as described in section 2.2.
Figure 4 shows the resulting light curves and illustrates the
location of the rectangular extraction apertures used. In Fig. 4, the total
brightness of the full OO~Ser reflection nebula, i.e., the largest extraction
aperture, is plotted on a different flux scale than the smaller sub-apertures.
It is clear from Fig.~4 that the overall brightness of the nebula reached
a relative minimum in 2002 or 2003, and has been slowly rising again after
that time, albeit with substantial local variations. 
This result was already indicated by two independent 
K$_s$-band photometric data points obtained in 2004 and 2006
by \citet {Kospal2007} (their Fig.~4). 

Up to now, the earliest published photometric data point in the
OO~Ser light curve had been the 1994 QUIRC image shown
in \citet{Hodapp1996}.
We have now re-analyzed the oldest sufficiently deep image of
OO~Ser available to us. 
This 1991 image obtained at the UH 2.2m telescope with the UH NICMOS3
camera data gives a 5$\sigma$ detection of the 
integrated flux of the OO~Ser nebula in pre-outburst
condition of K=16.8. 
This photometric value is about 2.5 magnitudes fainter than the
value of \textit{K}=14.3 mag for the
integrated flux measured in the same aperture on the 1994 QUIRC image. 
This demonstrates that by the time of the
1994 observations, the brightening process had already been under way.
We have also double-checked this result qualitatively by the artificial data experiment of adding
an appropriately intensity-scaled and smoothed version of an image
of OO~Ser in its bright state to the 1991 image. This 
experiment showed that OO~Ser would have been easily visible in
the 1991 image had it been at the same brightness as in 1994. 
Overall, this clearly shows that a significant brightening of OO~Ser began
before 1994 and maybe as early as 1991.

Our images in the 1--0 S(1) line of H$_2$ show shock excited emission in
the eastern lobe of the OO~Ser bipolar nebula. A thin arc of emission appears
to outline the edge of the eastern outflow cavity of OO~Ser in our H$-2$ 1-0 S(1)
narrow-band image in Fig. 1. The continuum-subtracted H$-2$ image in Fig. 2
shows that this arc ends in a shock front dominated by line emission. Still further to the east, several
individual shock fronts are visible whose proper motion (Fig.~1) traces them back to
OO~Ser. This proves that OO~Ser is an accreting source in the early evolutionary
state that is characterized by shock-excited H$_2$ emission. 
The H$_2$ shock fronts associated with OO~Ser have now been listed as
MHO 3245 in the catalog of molecular hydrogen shock fronts by \citet{Davis2010}.

The variable reflection nebulosity
extends to the west up to the apparent edge of the dense molecular cloud, as
judged from the density of background stars. 
To the east, the variable reflection
nebulosity extends a little beyond the 
fine line of S(1) emission just discussed,
but is, overall, much less extended than the reflection nebulosity in the western
direction. 
This morphology
strongly suggests that the western lobe, i.e., the
brighter reflection nebula, is slightly oriented towards the observer, 
while the eastern lobe with the H$_2$ shock fronts points away from the observer
and into denser parts of the Serpens NW molecular cloud so that reflection
nebulosity becomes un-observable. 

\subsection{EC~53 = V371~Ser}
The cometary nebula EC~53 shows many of the characteristics of a \textit{Class I}
young star seen in a nearly edge-on orientation of its accretion disk.
This object has been found by \citet{Hodapp1999} to be variable, with strong
indication of periodicity. 
We used the 
Period Analysis Software (Peranso) package 
written by T. Vanmunster, and several of
the algorithms implemented there for a period analysis of all our seeing-limited observations
of V371~Ser
in the past 20 years. The dominant period found by all algorithms is 543 days,
close to the period initially reported in \citet{Hodapp1999}. 
All the observations are plotted in a phase diagram with the 543 day period
solution and shown in Fig.~5.
Additional observations
are scheduled with the IRIS system
to better define the shape of the light curve.

Our \textit{K}-band adaptive optics images taken with
Keck/NIRC2 show this object in two very different phases
of its light curve. 
In 2007, the object was near minimum while in 2009, it was near maximum brightness.
The top panel of Fig. 6 shows the S(1) narrow-band image in 2007, near minimum brightness. 
The middle panel is a difference image of the S(1) line image minus a scaled \textit{K}
broad-band image, to bring out the H$_2$ jet near the center of the reflection nebula.
The bottom panel is a scaled difference image of the 2007 - 2009 data, showing the regions
varying in phase with the illuminating source in bright tones, while constant
(or un-phased) regions are shown in dark tone. The effectively constant
parts of the image include the companion object (EC~53~B), and the line emission
of the H$_2$ jet.
Further along the axis of this jet, outside of the field of our images
presented here, a terminal shock front was first
identified by \citet{Herbst1997} as their object S11, and is included
in the MHO catalog by \citet{Davis2010} as MHO~2221.
In the 2007 images, both in broadband K and in the narrow S(1) filter, 
it is clearly seen that that
the central star in EC~53 is a binary. The brighter component of this
binary system (EC~53 A) is the variable. In its brighter state, which was observed in 2009,
it so dominates the total flux that it makes the detection of the companion
star (EC~53 B) very difficult. A similar study of the uniformity of brightness changes
to that done on OO~Ser was done, but essentially resulted in the detection of
the nearly constant S(1) line emission component of the extended flux. Apart from
this, the change in brightness appeared uniform throughout the reflection nebula.

EC~53 must be a physical binary. 
The companion star is located at a projected distance of 296 mas (92 AU) from
the brighter component and has an apparent magnitude of \textit{K}=19.0. 
The nearest field star to EC~53 detected in our images is 14$\arcsec$ away.
With such a low density of unrelated stars, the probability that a 0.3$\arcsec$
companion is a chance alignment is $\approx$ 1/2000.
From the \textit{A$_V$} measurement of \citet{Eiroa1992}, and with the extinction
law of \citet{Rieke1985}, this gives an absolute \textit{K}-magnitude of 9.75 for 
the companion. 
The extinction value is based on the near-infrared color excess relative to normal stars
of the integrated light of the EC~53 nebula. 
The extinction towards the companion star may be significantly different, 
so the following discussion should be taken with some caution.

The Baraffe COND03 models \citep{Baraffe2003} for a 1 Myr star (the earliest models)
give \textit{M$_K$} = 9.73 for a 0.005 $M_{\sun}$ object, while the DUSTY models give a mass
between 0.004 and 0.005. The age of EC~53 is not well
determined, but based on its association with strong "cometary" nebulosity and
a molecular hydrogen jet, 
and the age of its immediate neighbors in Serpens NW, 
it is probably younger than 1 Myr. The uncertainties of the evolutionary
models are substantial, in particular at the very youngest ages, 
as pointed out by \citet{Baraffe2002}. 
We therefore take this mass as a very rough estimate
and conclude that the companion must be a future sub-stellar object, quite
likely of planetary mass.
At a projected distance of 92 AU, this low-mass companion is too far away from 
the main star to have been formed by core accretion \citep{Marley2007} at its
present distance. 
Also, 
less than 1 Myr is
not enough time to form giant planets by this mechanism. We therefore assume that 
the companion has formed at the same time as the main component by disk or core fragmentation. 

In their study of optical photometry of T Tauri stars, \citet{Herbst1994} list
three main causes for variability in these stars: rotational modulation of
dark star spots, modulation of bright accretion hot spots, and variable 
extinction due to eclipses by disks.
While variability is common in young stars, periodicity is mostly observed
on the timescale of days to a few weeks associated with stellar rotation.
For example, \citet{Carpenter2001} and \citet{Herbst2002} have studied variability
in the Orion Nebula at infrared and optical wavelengths, respectively.
While a large fraction of stars in their respective samples showed variability,
the most common type of variability was rotationally modulated variations of
spotted T Tauri stars with typical periods of order of several days to about
a week, and small peak-to-peak amplitudes of a fraction of a magnitude.
The short
time-span of most variability monitoring campaigns biases the results against
finding long-period variables. Even after considering this bias, 
longer period periodic variations in young stars appear to be very rare.
One example of longer period variability is WL4 in $\rho$~Oph, discovered by
\citet{Plavchan2008} in the 2MASS Calibration Point Source Working Database.
This star has a $\approx$0.4 mag amplitude in \textit{K$_s$} and a period of 130.87 days
caused by eclipses of each of two binary components by a circum-binary disk.
For WL4, \citet{Plavchan2008} specifically argue against a periodically driven
accretion model for this star.
On the other hand, \citet{Mathieu1997} interpreted the photometric variations
of DQ Tau (period 15.80 days) as evidence for ''variable accretion regulated
by the binary orbit''. Such variable accretion was also postulated by
\citet{Bouvier2003} as part
of the explanation of the variations (period 8.2 days) observed in AA Tau.
Among all the YSOs with established light curves, the long period, substantial
amplitude and association with ongoing accretion makes EC~53 = V371 Ser quite a unique
object.

Given that EC~53 is an outflow
source, we expect that most of its luminosity is derived from accretion
luminosity, and not from a, possibly pulsating, stellar atmosphere.
We therefore believe that the most likely explanation for the periodic
variations of EC~53 are periodic variations in the disk accretion rate,
modulated by the presence of a close companion star on an eccentric
orbit. This mechanism has been discussed in detail by \citet{Reipurth2000} as
an explanation for FUor outbursts and for giant, well-collimated outflows.
In this scenario,  an originally non-hierarchical multiple system evolves 
into a hierarchical system of a tightly bound pair and a third component at
a much larger distance that may ultimately become unbound. This model has
recently been substantiated by detailed orbital simulation by \citet{Reipurth2010}.
In a tight binary system, the main reservoir of material available for accretion
lies outside of the binary orbit, explaining the long-wavelength spectral energy distribution peak for
young accreting binaries. It has been shown by \citet{Artymowicz1996} that
streams of material can flow through the binary orbits into the immediate vicinity 
of either of the two close binary components, from where they can be accreted onto the stars by
magnetospheric accretion. These models have been detailed out by recent
numerical simulations by \citet{deValBorro2011}. Even though the relationship
between binary orbit and accretion streams is not trivial, the model calculations
show that in eccentric orbits, the arrival of the accretion streams at the star
roughly coincides with the periastron phase of the binary orbit.
In our case of EC~53, the observed companion (EC~53 B) at 296 mas projected separation is
the distant third component, while the two components responsible for the 543 day
period of light variation form the closely bound system EC~53 A.
The formation of a tightly bound binary may lead to the release of the third component
of the original non-hierarchical system, 
in the case of EC~53 
a free-floating object of maybe only planetary mass.

So far, no repetitive FUor events have been found. Some of the smaller EXor outbursts 
have been found to be repetitive, for example the prototypical EX Lupi \citep{Aspin2010}
or V1647 Ori \citep{Aspin2006}, but there are no cases with a clearly established periodicity.
EC~53 may therefore represent the short-period extension of the FUor and EXor
phenomenon.

\section{CONCLUSIONS}
We have presented the results of Keck/NIRC2 laser adaptive optics imaging
of three young, accreting objects in the Serpens NW region.

Much of the shock-excited H$_2$ emission between the eruptive variable OO~Ser
and the irregular variable EC~37 = V370 Ser originates in the latter. The proper motion
pattern and the relatively low extinction to this object suggest that the line of sight 
towards this object is closely aligned
with the outflow axis. 

We have identified small patches of H$_2$ emission east of OO~Ser, associated with
the bipolar nebula centered on this deeply embedded outburst star. These shock fronts demonstrate that
OO Ser has ongoing outflow activity and is therefore still accreting mass. 
Based on images spread over many years, we show
variations in the light distribution of the western lobe of the OO~Ser bipolar nebula that
are similar to other cases of "shadow play" in young reflection nebulae, indicative of
spatially and temporarily variable extinction near the star in the line of sight of the
reflected light. After the 1995 outburst of OO~Ser, both the central source and the 
integrated light of the surrounding bipolar reflection nebula faded until about
2003. After this time, the overall decline has stopped. Within the range of the spatial
variations in brightness distribution observed before, some areas of the reflection nebula
have brightened, while others have continued to fade. 

The variable cometary nebula EC~53 = V371 Ser has an H$_2$ jet, indicating ongoing
outflow activity. The earlier suggestion that the brightness variations of this object
are periodic has been confirmed by many additional observations and a period of 543 d is
now well established. V371 Ser has a faint, non-variable companion at a projected distance
of 92~AU that is probably a very low mass future brown dwarf or even an object of only planetary mass.
The variable cometary nebula EC~53 is being interpreted in the framework of the \citet{Reipurth2000} model
whereby the ejection of a third component from a previously non-hierarchical system
leads to the formation of a tight binary on an initially eccentric orbit that then
imposes its orbital period on the on-going disk accretion, leading to a periodicity
in the accretion luminosity.

\acknowledgments

Some of the data presented herein were obtained at the W.M. Keck Observatory,   
which is operated as a scientific partnership among the 
California Institute of Technology,
the University of California and NASA.
The Observatory was made possible by the generous financial support of the W.M. Keck Foundation.

The 2.2m telescope on Mauna Kea is operated by the University of Hawaii.

The United Kingdom Infrared Telescope is operated by the Joint Astronomy Centre on behalf of the
Science and Technology Facilities Council of the U.K.

This paper is based on observations obtained with WIRCAM, a joint project of CFHT, Taiwan, Korea, Canada, France,
at the Canada France-Hawaii Telescope (CFHT), which is operated by the National Research Council (NRC)
of Canada, the Institute National des Sciences de l'Univers of the Centre National de la 
Recherche Scientifique of France, and the University of Hawaii.

This publication makes use of data products from the Two Micron All Sky Survey, which is
a joint project of the University of Massachusetts and the Infrared Processing and Analysis Center/
California Institute of Technology, funded by the National Aeronautics and Space Administration
and the National Science Foundation.

The Infrared Imaging Survey (IRIS) is a joint project of the University of Hawaii, the Ruhr Universit\"{a}t in
Bochum, Germany, and the Universidad Catolica del Norte in Antofagasta, Chile. 
We thank L.-S. Buda, M. D{\"o}rr, H. Drass, and V. Hoffmeister for their help with operating the IRIS telescope. 
We thank B. Reipurth for helpful discussions and a careful reading of the manuscript.
We thank the referee, Chris J. Davis, for helpful comments.

Development of the IRIS camera was supported by the National Science Foundation under grant AST~0704954.

\clearpage
\begin{deluxetable}{cccccccc}
\tabletypesize{\scriptsize}
\tablecaption{Log of Keck AO Observations}
\tablewidth{0pt}
\tablehead{
\colhead{Date (UT)} & \colhead{JD} & \colhead{Filter} & \colhead{t} & \colhead{nreads} & \colhead{coadds} & \colhead{frames used} & \colhead{FWHM [mas]}
}
\startdata
2007-06-10 & 2454262 & H$_2$ & 30s &  8 &  1 & 125 & 88\\
2007-07-09 & 2454291 & K  &  2s &  8 & 45 & 25 & 123\\
2009-06-20 & 2455003 & H$_2$ & 60s & 16 &  1 & 82 & 114\\
2009-06-29 & 2455012 & K  &  2s &  8 & 45 & 15 & 97\\
\enddata
\end{deluxetable}

\clearpage
\begin{deluxetable}{cccc}
\tabletypesize{\scriptsize}
\tablecaption{Log of Observations}
\tablewidth{0pt}
\tablehead{
\colhead{Date} & \colhead{JD} & \colhead{Instrument} & \colhead{Filter}
}
\startdata
5  Aug 1991 & 2448474 & UH NICMOS3 & K'\\
16 Aug 1994 & 2449581 & UH QUIRC & K\\
13 Jul 1995 & 2449912 & UH QUIRC & K\\
12 Oct 1995 & 2450003 & UH QUIRC & K\\
6  Feb 1996 & 2450120 & UH QUIRC & K\\
28 Jul 1996 & 2450293 & UH QUIRC & K\\
3  Oct 1996 & 2450360 & UH QUIRC & K\\
25 Apr 1997 & 2450564 & UH QUIRC & K\\
17 Aug 1997 & 2450678 & UH QUIRC & K\\
5  Feb 1998 & 2450850 & UH QUIRC & K\\
30 Sep 1998 & 2451087 & UH QUIRC & K\\
18 Aug 2000 & 2451775 & UH SIRIUS & K$_s$\\
1  Aug 2002 & 2452488 & UH QUIRC & K$_{MKO}$\\
10 Jun 2003 & 2452801 & UH QUIRC & K$_{MKO}$\\
31 Jul 2004 & 2453218 & UH QUIRC & K$_{MKO}$\\
21 May 2008 & 2454608 & UKIRT WFCAM & K$_{MKO}$\\
17 Aug 2010 & 2455426 & IRIS & K$_s$\\
28 Aug 2010 & 2455436 & IRIS & K$_s$\\
14 Sep 2010 & 2455454 & IRIS & K$_s$\\
1 Oct 2010 & 2455471 & IRIS & K$_s$\\
20 Oct 2010 & 2455490 & IRIS & K$_s$\\
20 Mar 2011 & 2455641 & IRIS & K$_s$\\
\enddata
\end{deluxetable}

\clearpage
\begin{figure}
\figurenum{1}
\includegraphics[scale=0.8,angle=-90]{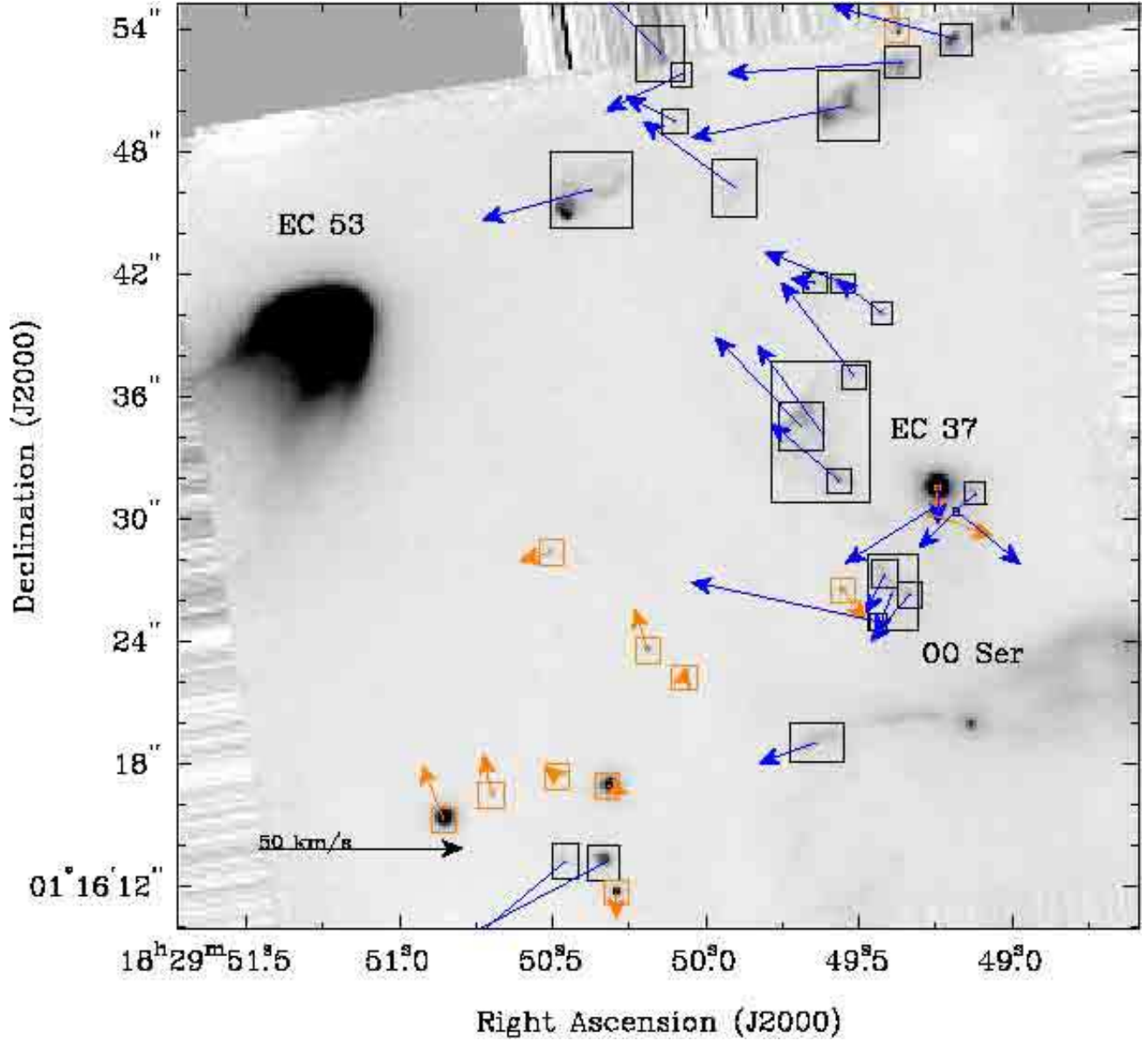}
\caption{
Proper motion vectors measured on the Keck/NIRC2 2.12 $\mu$m images over
the time interval 2007 - 2009. The underlying image is the 2009 Keck NIRC2 LGSAO 2.12 $\mu$m image.
Amber (light grey) arrows indicate the proper motion of stars in the field, and are largely an
indication of the uncertainties of the astrometric reference that is based on the
ensemble of these stars. Blue (dark grey) arrows indicate the measured proper motions of 
H$_2$ shock fronts.
}
\end{figure}

\clearpage
\begin{figure}
\figurenum{2}
\includegraphics[scale=0.9,angle=0]{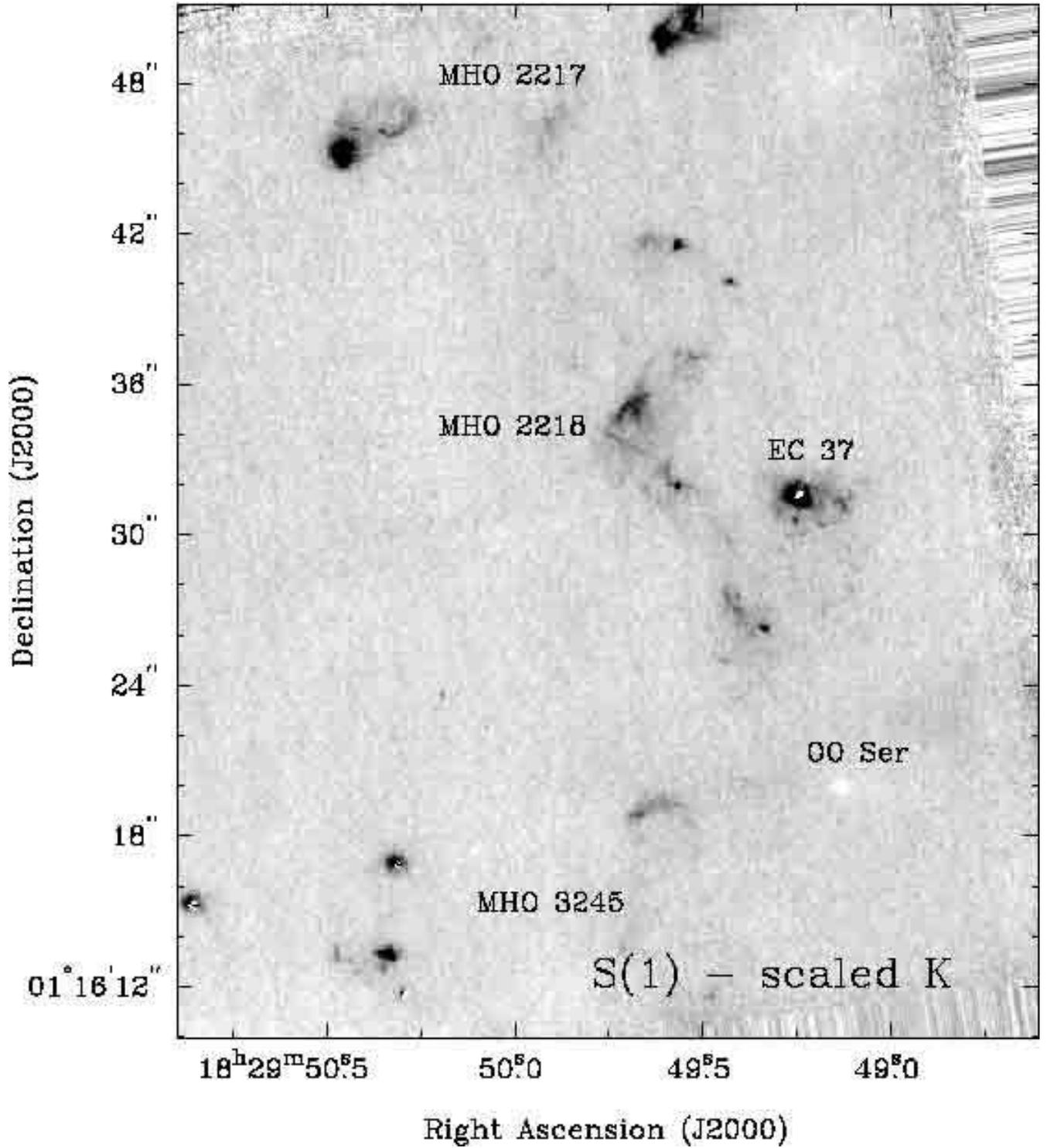}
\caption{
Continuum subtracted v=1-0 S(1) image of the shock fronts associated
with EC~37 and OO~Ser based on the 2009 Keck data. The broad-band (continuum) image was scaled
to subtract out the flux of the OO~Ser reflection nebula. The reddest
object in the frame, the central condensed area of this reflection 
nebula, therefore appears white (over-subtracted) while all other, less
reddened continuum objects, appear black.
}
\end{figure}

\clearpage
\begin{figure}
\figurenum{3}
\includegraphics[scale=0.9,angle=0]{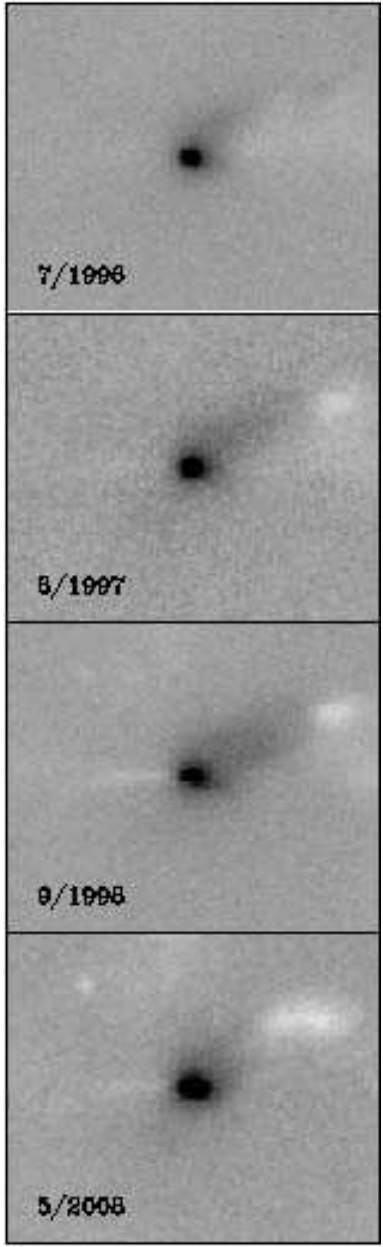}
\caption{
Scaled difference images based on seeing-limited \textit{K}-band images of the OO Ser reflection
nebulosity between 1996 and 2008. The field of view is approximately $23\arcsec\times19\arcsec$. 
All images used were first scaled by the
integrated signal of the reflection nebula to account for the overall brightness
variations of the object. For each of the months and years indicated (for precice dates see
Table 2), the difference image is the flux-scaled 1995 peak brightness image minus the
image taken at the given date.
In this scaling, white tone indicates slower fading than average, while dark tone
indicates faster than average fading.
}
\end{figure}

\clearpage
\begin{figure}
\figurenum{4}
\includegraphics[scale=0.75,angle=-90]{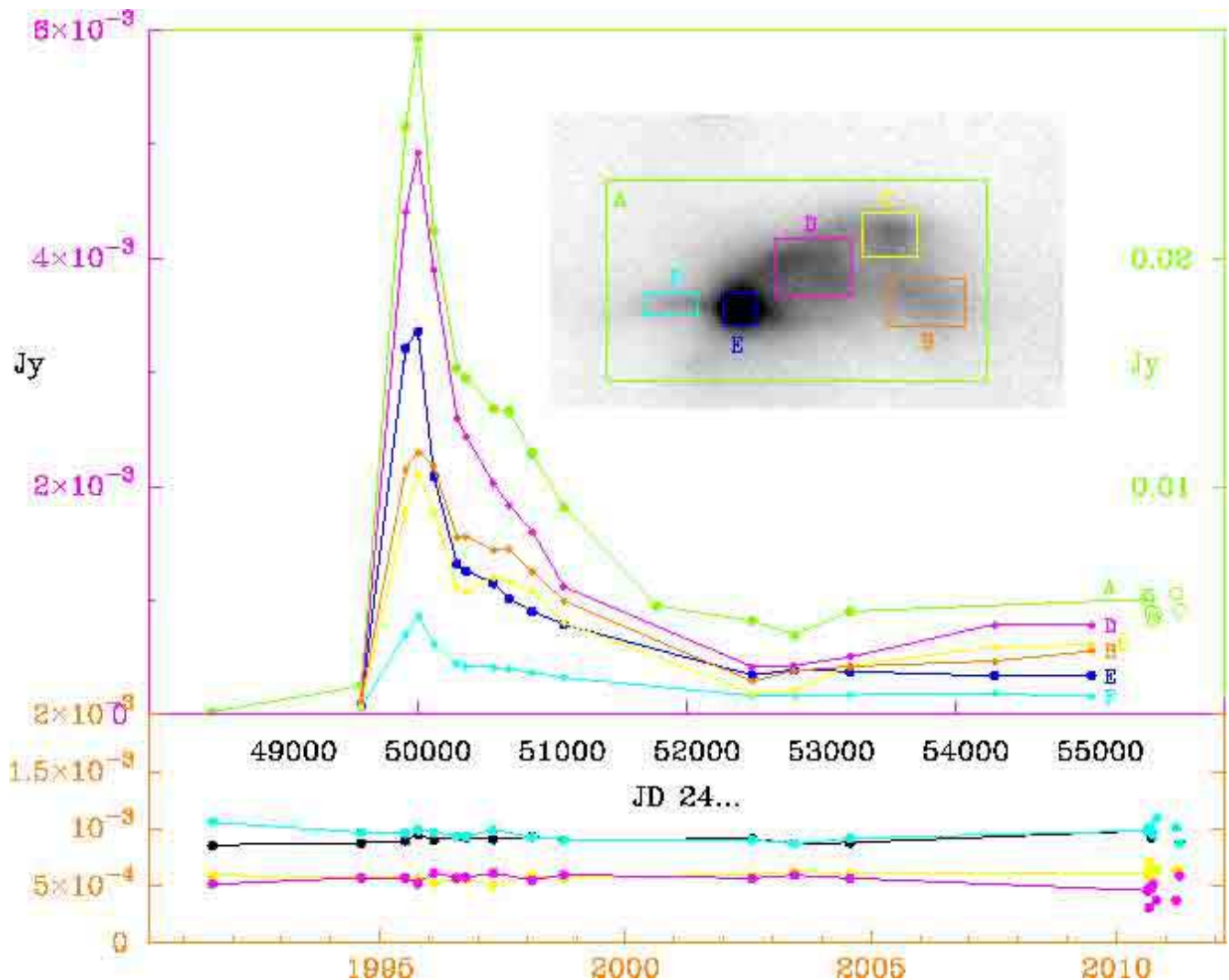}
\caption{
Light curve of the integrated \textit{K}-band light from OO Ser and from individual regions
of the reflection nebulosity. This photometry is based on all the
observations listed in Tables 1 and 2.
The top (green) light curve represents the integral flux over the whole
reflection nebula. The flux calibration for this curve is to the right
side of the figure. The other light curves are for the smaller sub-regions
of the reflection nebula, and they refer to the flux scale on the left, which
is a factor of 5 smaller than the right scale.
The lower panel of the figure shows, on the same flux scale as for the sub-apertures, the measurements
of the four nominally constant reference stars, as an indication of typical
total measurement errors.
The horizontal axis gives the time both in abbreviated Julian dates and calendar years.
}
\end{figure}

\clearpage
\begin{figure}
\figurenum{5}
\includegraphics[scale=0.9,angle=0]{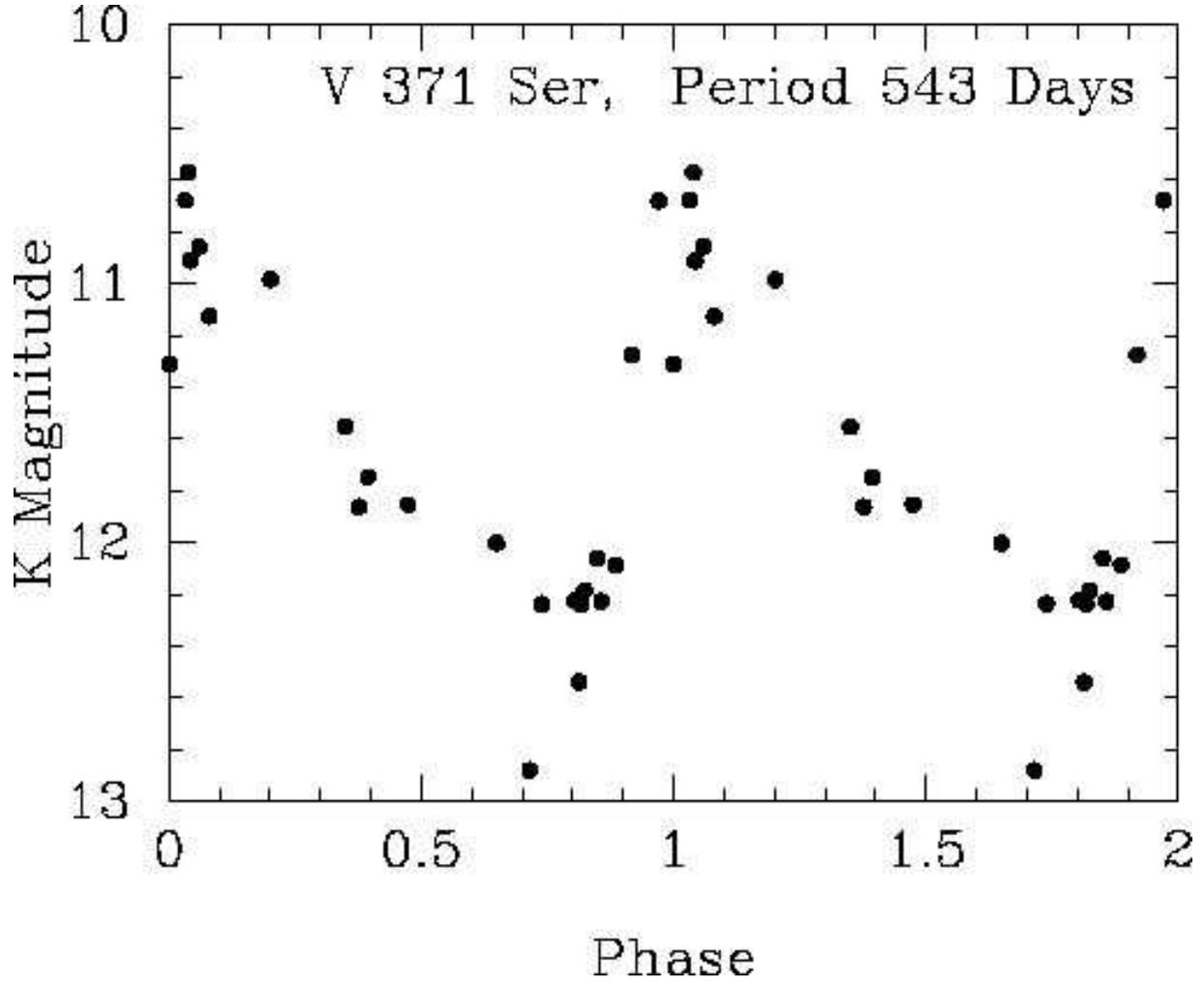}
\caption{
Phase diagram of the \textit{K}-band variations of EC~53.
The light curve shows a rapid rise and slower decline from the
maximum.
}
\end{figure}

\clearpage
\begin{figure}
\figurenum{6}
\includegraphics[scale=0.75,angle=0]{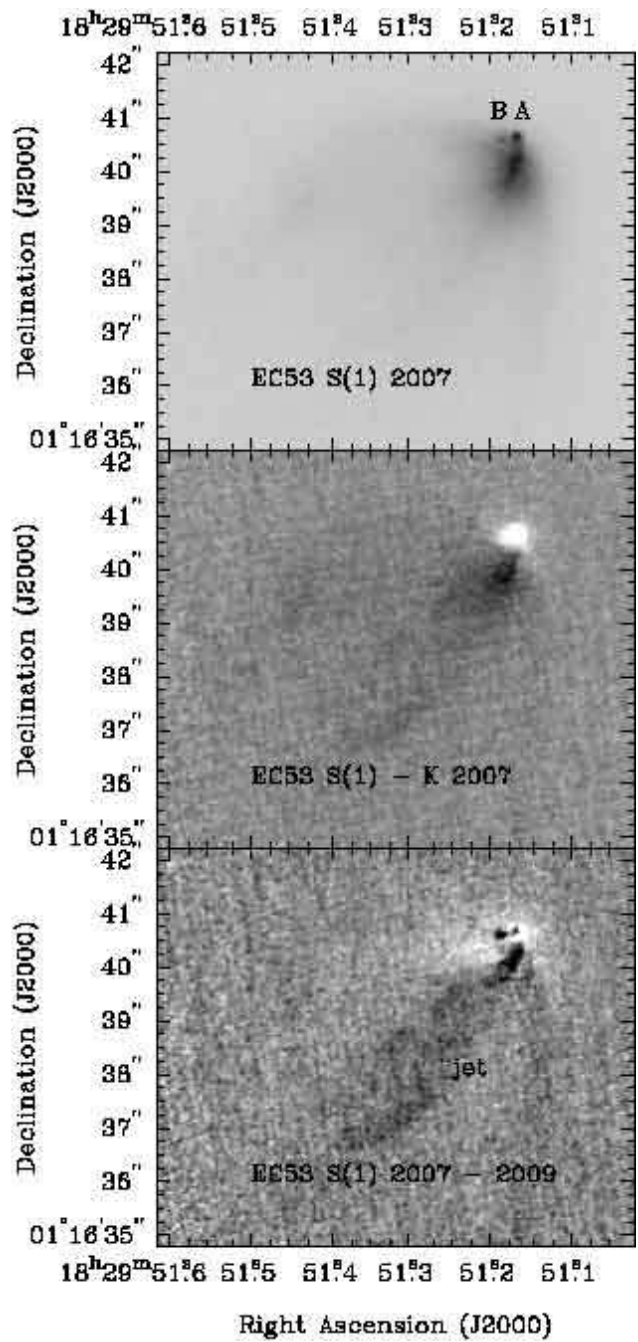}
\caption{
Top panel: Keck NIRC2 LGSAO image in the 2.12$\mu$m S(1) filter of
V371 Ser = EC~53 taken near minimal light in 2007. 
Middle panel: Difference image of the S(1) line image and a \textit{K} broad-band
image. The broad band image was scaled to that the reddest object in the
field, the variable source V371 Ser A, appears white, the companion
V371 Ser B is subtracted away, and regions with strong line emission
appear dark. The line-dominated flux extends to the south-east from
V371 Ser A, while the overall reflection nebula (top panel) extends
straight to the south.
Lower panel: Difference of 2.12$\mu$m S(1) line images taken in
2007 (near minimum) and 2009 (close to maximum) of V371 Ser.
The image was scaled to show regions varying strongly and in phase
with V371 Ser in white, while regions of constant flux are dark.
The companion V371 Ser B is visible as a black dot, indicating that
it does not vary in phase with component A, i.e., is not a reflection
nebula.
}
\end{figure}


\clearpage

\begin{thebibliography}{}

\bibitem[Adams, Lada, \& Shu (1987)]{Adams1987}
Adams, F. C., Lada, C. J., \& Shu, F. H. 1987, \apj, 312, 788

\bibitem[Artymowicz \& Lubow (1996)]{Artymowicz1996}
Artymowicz, P. \& Lubow, S. H. 1996, \apj, 467, L77

\bibitem[Aspin et al.(2006)]{Aspin2006}
Aspin, C., Barbieri, C., Boschi, F., Di Mille, F., Rampazzi, F, Reipurth, B, \& Tsvetkov, M. 2006, \aap, 132, 1298

\bibitem[Aspin et al.(2010)]{Aspin2010}
Aspin, C., Reipurth, B., Herczeg, G. J., Capak, P. 2010, \apjl, 719, 50

\bibitem[Baraffe et al.(2003)]{Baraffe2003}
Baraffe, I., Chabrier, G., Barman, T. S., Allard, F., \& Hauschildt, P. H. 2003, \aap, 402, 701

\bibitem[Baraffe et al.(2002)]{Baraffe2002}
Baraffe, I., Chabrier, G., Allard, F., \& Hauschildt, P. H. 2002, \aap, 382, 563

\bibitem[Bouvier et al.(2003)]{Bouvier2003}
Bouvier, J., Grankin, K. N., Alencar, S. H. P., Dougados, C., Fernandez, M.,
Basri, G., Batalha, C., E. Guenther, Ibrahimov, M. A., Magakian, T. Y.,
Melnikov, S. Y., Petrov, P. P., Rud, M. V., \& Zapatero Osorio 2003, \aap, 409, 169

\bibitem[Carpenter, Hillenbrand, \& Skrutskie (2001)]{Carpenter2001}
Carpenter, J. M., Hillenbrand, L. A., \& Skrutskie, M. F. 2001, \aj, 121, 3160

\bibitem[Cohen, Wheaton, \& Megeath (2003)]{Cohen2003}
Cohen, M., Wheaton, Wm. A., \& Megeath, S. T. 2003, \aj, 126, 1090

\bibitem[Connelley, Hodapp, \& Fuller (2009)]{Connelley2009}
Connelley, M. S., Hodapp, K. W., \& Fuller, G. A. 2009, \aj, 237, 3494

\bibitem[Davis et al.(2010)]{Davis2010}
Davis, C. J., Gell, R., Khanzadyan, T., Smith, M. D., \& Jenness, T.
2010, \aap, 511, 24

\bibitem[de Lara, Chavarria-K. \& Lopez-Molina (1991)]{deLara1991}
de Lara, E., Chavarria-K., C., \& Lopez-Molina, G. 1991, \aap, 243, 139

\bibitem[de Val-Borro et al.(2011)]{deValBorro2011}
de Val-Borro, M., Gahm, G. F., Stempels, H. C., \& Peplinski, A. 2011,
\mnras, 413, 2679

\bibitem[Eiroa \& Casali(1992)]{Eiroa1992}
Eiroa, C., \& Casali, M. M. 1992, \aap, 262, 468

\bibitem[Evans et al.(2009)]{Evans2009} 
Evans, N.~J., et al. 2009, \apjs, 181, 321 

\bibitem[Herbig (1977)]{Herbig1977}
Herbig, G. H. 1977, \apj, 217, 693

\bibitem[Herbst, Beckwith, \& Robberto(1997)]{Herbst1997}
Herbst, T. M., Beckwith, S. V. W., \& Robberto, M. 1997,
\apj, 486, L59

\bibitem[Herbst, Herbst \& Grossman (1994)]{Herbst1994}
Herbst, W., Herbst, D. K., \& Grossman, E. J. 1994, \aj, 108, 1906

\bibitem[Herbst et al.(2002)]{Herbst2002}
Herbst, W., Bailer-Jones, C. A. L., Mundt, R., Meisenheimer, K., \& Wackermann, R. 2002, \aap, 396, 513

\bibitem[Hodapp, Rayner, \& Irwin (1992)]{Hodapp1992}
Hodapp, K. W., Rayner, J. , \& Irwin, E. 1992, \pasp, 104, 441

\bibitem[Hodapp et al.(1996a)]{Hodapp1996}
Hodapp, K.-W., Hora, J. L., Rayner, J. T., Pickles, A. J. \& Ladd, E. F. 1996,
\apj, 468, 861

\bibitem[Hodapp et al.(1996b)]{Hodapp1996QUIRC}
Hodapp, K.-W. et al. 1996, New Astronomy 1, 177

\bibitem[Hodapp(1999)]{Hodapp1999}
Hodapp, K.-W. 1999, \aj, 118, 1338

\bibitem[Hodapp \& Bressert (2009)]{Hodapp2009}
Hodapp, K. W., \& Bressert, E. 2009, \aj, 137, 3501

\bibitem[Hodapp et al.(2010)]{Hodapp2010}
Hodapp, K. W., Chini, R., Reipurth, B., Murphy, M., Lemke, R.,
Watermann, R., Jacobson, S., Bischoff, K., Chonis, T., Dement, K.,
Terrien, R., \& Provence, S. 2010, Proc. SPIE 7735-45.

\bibitem[Hubble (1916)]{Hubble1916}
Hubble, E. 1916, \apj, 44, 190

\bibitem[Johnson (1966)]{Johnson1966}
Johnson, H. L. 1966, ARAA, 4, 193

\bibitem[Joy (1945)]{Joy1945}
Joy, A. H. 1945, \apj, 102, 168

\bibitem[K{\'o}sp{\'a}l et al.(2007)]{Kospal2007} 
K{\'o}sp{\'a}l, {\'A}., {\'A}brah{\'a}m, P., Prusti, T., 
Acosta-Pulido, J., Hony, S., Mo{\'o}r, A., \& Siebenmorgen, R.\ 2007, \aap, 470, 211 

\bibitem[Marley et al.(2007)]{Marley2007}
Marley, M. S., Fortney, J. J., Hubickyj, O., Bodenheimer, P.,
\& Lissauer, J. J. 2007, \apj, 655, 541

\bibitem[Mathieu et al.(1997)]{Mathieu1997}
Mathieu, R. D., Stassun, K., Basri, G., Jensen, E. L. N., Johns-Krull, C. M.,
Valenti, J. A., \& Hartmann, L. W. 1997, \aj, 113, 1841

\bibitem[Nakajima et al.(2002)]{Nakajima2002}
Nakajima, Y., Tamura, M., Nagashima, C., Nagayama, T., Baba, D., Sugitani, K.,
Nakaya, H., Nagata, T., Kato, D., Kurita, M., Sato, S., Jiang, Z., Ita, Y.,
Tanabe, T., Nakata, Y., Naoi, T., Oasa, Y. 2002,
Proc. SPIE 4836, 29-34

\bibitem[Plavchan et al.(2008)]{Plavchan2008}
Plavchan, P., Gee, A. H., Stapelfeldt, K., \& Becker, A. 2008, \apj, 684, L37

\bibitem[Reipurth (2000)]{Reipurth2000}
Reipurth, B. 2000, \aj, 120, 3177

\bibitem[Reipurth et al.(2010)]{Reipurth2010}
Reipurth, B., Mikkola, S., Connelley, M., \& Valtonen, M. 2010,
\apjl, 725, L56

\bibitem[Rieke \& Lebofsky (1985)]{Rieke1985}
Rieke, G. H. \& Lebofky, M. J. 1985, \apj, 288, 618


\bibitem[Skrutskie et al.(2006)]{skr06}
Skrutskie, M. F., Cutri, R. M., Stiening, R., Weinberg, M. D., Schneider, S.,
Carpenter, J. M., Beichman, C., Capps, R., Chester, T., Elias, J., Huchra, J.,
Liebert, J., Lonsdale, C., Monet, D. G., Price, S., Seitzer, P., Jarrett, T.,
Kirkpatrick, J. D., Gizis, J., Howard, E., Evans, T., Fowler, J., Fullmer, L., 
Hurt, R., Light, R., Kopan, E. L., Marsh, K. A., McCallon, H. L., Tam, R.,
Van Dyk, S., \& Wheelock, S. 2006, \aj, 131, 1163

\bibitem[Testi \& Sargent (1998)]{Testi1998}
Testi, L., \& Sargent, A. I. 1998, \apj, 508, L91

\bibitem[Tokunaga, Simons, \& Vacca (2002)]{Tokunaga2002}
Tokunaga, A. T., Simons, D. A., \& Vacca, W. D. 2001, \pasp, 114, 180

\bibitem[Wainscoat \& Cowie (1992)]{Wainscoat1992}
Wainscoat, R. J. \& Cowie, L. L. 1992, \aj, 103, 332

\bibitem[Wizinowich et al.(2006)]{Wizinowich2006}
Wizonowich, P. L., Le Mignant, D., Bouchez, A. H., Campbell, R. D., Chin, J. C. Y.,
Contos, A. R., Van Dam, M. A., Hartman, S. K., Johansson, E. M., Lafon, R. E.,
Lewis, H., Stomski, P. J., \& Summers, D. M. 2006, \pasp, 118, 297

\end{thebibliography}
\end{document}